\documentclass[12pt]{article}

\usepackage{amsthm,amsmath,amssymb,bbm,bm}
\usepackage{multirow}
\usepackage[pdftex]{graphicx}
\usepackage{subfigure}
\usepackage{wrapfig}
\usepackage{CJK}
\usepackage{array}
\usepackage{url}
\usepackage{booktabs}
\usepackage{algorithm}
\usepackage{algorithmic}
\usepackage{placeins}
\usepackage{mathtools}
\usepackage{romannum}
\usepackage{mathrsfs}
\usepackage{dsfont}
\usepackage{natbib}
\usepackage{relsize}
\usepackage{rotating}
\usepackage{enumitem}
\usepackage{setspace}
\usepackage{minitoc}
\usepackage{etoc}
\usepackage{listings}
\usepackage{hyperref}
\usepackage{authblk}

\usepackage[usenames,dvipsnames,svgnames,table]{xcolor}

\makeatletter
\newcommand{\vast}{\bBigg@{3}}
\makeatother

\usepackage{chngcntr}
\counterwithin{equation}{section}
\definecolor{ao(english)}{rgb}{0.0, 0.5, 0.0}
	
\theoremstyle{definition}

\newtheorem{thm}{Theorem}

\counterwithin{thm}{section}
\newtheorem{rmk}{Remark}[thm]

\counterwithin{rmk}{section}
\counterwithin{proposition}{section}

\counterwithin{corollary}{section}

\def\##1\#{\begin{align}#1\end{align}}
\def\$#1\${\begin{align*}#1\end{align*}}

\def\beq#1\eeq{\begin{equation}#1\end{equation}}
\def\baa#1\eaa{\begin{eqnarray}#1\end{eqnarray}}
\def\bal#1\eal{\begin{align}#1\end{align}}

\DeclareMathOperator*{\argmax}{arg\,max}

\usepackage[margin=0.85in]{geometry}

\makeatletter
\def\@fnsymbol#1{\ensuremath{\ifcase#1\or \mathparagraph \or  \dagger \or \ddagger\or
   \dagger\or \mathparagraph\or \|\or **\or \dagger\dagger
   \or \ddagger\ddagger \else\@ctrerr\fi}}
    \makeatother

\begin{document}

\title{\bf Policy Learning for Individualized Treatment Regimes on Infinite Time Horizon}
\author[1]{Wenzhuo Zhou\thanks{\href{mailto:wenzhuz3@uci.edu}{wenzhuz3@uci.edu}}}
\author[2]{Yuhan Li\thanks{\href{mailto:yuhanli8@illinois.edu}{yuhanli8@illinois.edu}}}
\author[2]{Ruoqing Zhu\thanks{\href{mailto:rqzhu@illinois.edu}{rqzhu@illinois.edu}}}
\affil[1]{Department of Statistics, University of California Irvine}
\affil[2]{Department of Statistics, University of Illinois Urbana-Champaign}
 \date{}
 \maketitle


\setcounter{page}{1}
\pagenumbering{arabic}

\begin{abstract}
With the recent advancements of technology in facilitating real-time monitoring and data collection, ``just-in-time'' interventions can be delivered via mobile devices to achieve both real-time and long-term management and control. Reinforcement learning formalizes such mobile interventions as a sequence of decision rules and assigns treatment arms based on the user's status at each decision point. In practice,  real applications concern a large number of decision points beyond the time horizon of the currently collected data. This usually refers to reinforcement learning in the infinite horizon setting, which becomes much more challenging. This article provides a selective overview of some statistical methodologies on this topic. We discuss their modeling framework, generalizability, and interpretability and provide some use case examples. Some future research directions are discussed in the end. 
\end{abstract}

\section{Background and Challenges}

The main purpose of this chapter is to review the statistical methods developed under the infinite horizon setting. Under a finite horizon setting, the optimal treatment regimes can only be estimated within a fixed period with limited decision points \citep{chakraborty2014dynamic}. However, with the rapid development of sensor technology and the increasing popularity of various wearable devices, a tremendous amount of personal health data are available to healthcare providers and researchers \citep{steinhubl2015emerging}. For example, such devices may record personal health information over an extremely long period \citep{silva2015mobile}, making it possible to construct personalized treatment plans for managing chronic diseases and other health issues over an infinite horizon with a large number of decision points. Hence, research on mobile health (mHealth) has emerged \citep{rehg2017mobile} in recent years. The focus is to use mobile technologies to improve healthcare quality regardless of location and time. Research and practice in this field face challenges that span many disciplines and fields, including but not limited to engineering, computer science, statistics, bioinformatics, medicine, health policy, and privacy. 
  
To date, mHealth has been used extensively in managing stress, treating depression, and other chronic diseases such as diabetes and cardiovascular diseases. It plays an essential role in assisting healthcare providers to better monitor and treat patients. For instance, the AutoSense suite can detect physiological signals (ECG, respiration, etc.) at a high frequency (approximately 5 million samples per day) and allow machine learning models to convert the raw data into biomarkers of health, behavior, and environment with smartphone software \citep{sarker2017markers}. Interventions can be delivered to users promptly if these biomarkers suggest an excessive risk of stress or other health conditions. However, certain limitations exist in the current development of mHealth. Most existing methods consider only the finite horizon setting and are not directly applicable to the infinite horizon setting. Additionally, such methods may require strong assumptions that are difficult to justify or validate in real-life \citep{kosorok2019precision}. Recent literature mainly borrows tools developed in reinforcement learning \citep{sutton2018reinforcement, ertefaie2018constructing, luckett2020estimating, shi2020statistical, zhou2022estimating, liao2022batch, shi2022statistically},  to address the infinite horizon problem, although certain limitations still exist, such as data coverage, function approximation, Markovian, and many others \citep{fujimoto2019off, cheung2020reinforcement, levine2020offline, uehara2022review}. 

This book chapter mainly reviews some statistical methods for estimating the optimal treatment regimes under an infinite horizon setting. We will introduce several popular approaches and related concepts, such as Q-learning, temporal difference learning, and the residual gradient algorithm. We also discuss some developments motivated by the limitations of these classical methods, including the greedy gradient Q-learning, V-learning\footnote{We choose greedy gradient Q-learning and V-learning methods for detailed illustration in this paper as they are one of the most pioneering works in the statistics community for DTR in infinite horizon settings during these years. Most recently, there have been many interesting and important works emerged within the community. Due to the page limitation, we make apologize in advance that we cannot deeply review all of them in this paper.}, and a newly proposed temporal consistent learning framework in statistical community.

Our review is selective in the sense that we focus mostly on methods that can be directly applied to an already collected dataset from an observational study. Hence, they are offline and off-policy. Extensions to online settings are briefly discussed. Some examples are used to demonstrate existing applications. We conclude this chapter with a discussion of open issues on this topic.

\section{Preliminaries}

\subsection{A Motivating Example} \label{motex}

\begin{figure}[h]
\centering
\includegraphics[width=13cm]{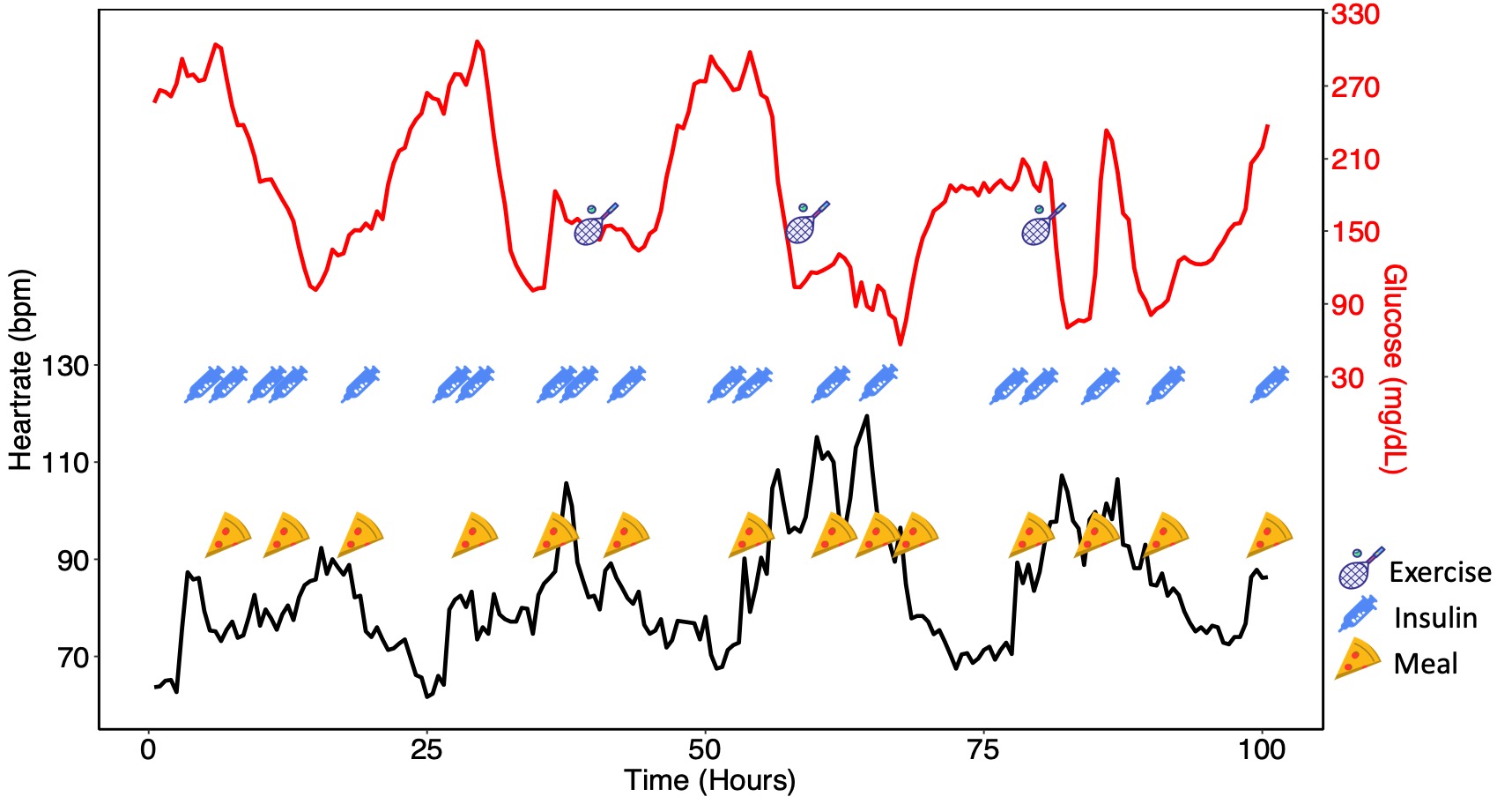}
\caption{OhioT1DM Data: A longitudinal observation of a patient}\label{fig:OhioT1DM}
\end{figure}

Glucose management for patients with diabetes is one of the most widely used applications of mHealth. By continuously monitoring the glucose level, food intake, and physiological information, a series of Just-in-time interventions, such as insulin injection, can be delivered to patients to improve long-term benefits \citep{muralidharan2017mobile}. An application example is the OhioT1DM study \citep{marling2020ohiot1dm}, which contains 12 patients with Type 1 Diabetes. This dataset includes a continuous glucose monitoring (CGM) of blood glucose levels every 5 minutes over eight weeks for each patient, along with self-reported times of meal, exercise, sleep, work, and stress. The time points of insulin doses are also recorded. Figure \ref{fig:OhioT1DM} provides a snapshot of the fluctuation of glucose level, insulin injections, meals, exercise, and heart rate of a patient during a 100 hour time interval. 

Patients may respond differently to insulin injections due to differences in age, gender, and other personal activity and health measurements. We want to develop a framework to understand the dynamic interactions between insulin injection and glucose level. This also allows us to determine a personalized treatment policy that could best control the glucose level of each individual. 

\subsection{Notation}\label{note}

To better formulate this problem, we introduce some important notations and concepts. Denote $S^t\in \mathbb{R}^p$ as the patient information collected at time $t$, which may include all related covariates such as age, gender, and lab results. The values of these covariates are likely to change between time points, reflecting the patient's instantaneous condition. We denote $A^t\in \mathcal{A}$ as the treatment assigned to the patient at time $t$ based on the information $S^t$, which may include the drug choice and dosing level. Here, $\mathcal{A}$ could be either a discrete or continuous space for possible treatments. Hence, the collected dataset is  $\{S_i^0,A_i^0,S_i^1, \ldots ,S_i^{T},A_i^{T},S_i^{T+1}\}^n_{i=0}$, which compromises $n$ i.i.d. trajectories with $T+1$ time points of treatment decisions. We further assume an observed immediate reward $R^t$ that may depend on $S^{t+1},A^t,S^t$, measured after choosing treatment $A^t$ at time $t$. Specifically, it measures how well the patient reacts to treatment $A^t$ as the covariates change from $S^t$ to $S^{t+1}$. Typically, a larger value of reward means a better treatment outcome.

In the previous motivating example, $S^t$ can be the patient's glucose level, physical activity, age, gender, dietary intake, and other related information contained in the dataset. $A^t$ is the indicator or dosing of insulin injection, and $R$ can be defined as a measure of closeness of the patient's glucose level from the optimal glucose level range. Our goal is to maximize $R$ for each individual by learning the optimal strategy of insulin injection. We first introduce several essential concepts. 

\textbf{Definition 1: (Markov Property)} Suppose $s^t\in \mathcal{S}$ denotes the observed state at time $t$, $a^t \in \mathcal{A}$ denotes the action taken at time $t$, then under the Markov property :
$$P(S^{t+1} = s^{t+1}| S^t = s^t, \ldots, S^1 = s^1, A^t = a^t, \ldots , A^0 = a^0)=P(S^{t+1} = s^{t+1}|S^t = s^t, A^t = a^t).$$
The Markov property suggests that the current state and action are sufficient to predict the next state.

\textbf{Definition 2: (Markov Decision Process, MDP)} A Markov decision process (MDP) is an environment in which all states satisfy the Markov property, and a decision maker can interact with the environment using actions. Specifically, it is a tuple $<\mathcal{S},\mathcal{A},\mathbf{P},R>$ where
 \begin{itemize}
  \item $\mathcal{S}$ is a set of states called state space, which may be finite or infinite;
  \item $\mathcal{A}$ is a set of actions called action space, which may be finite or infinite;
  \item $\mathbf{P}(s^{t+1}, a^t, s^t)=P(S^{t+1}=s^{t+1}|S^t=s^t,A^t=a^t)$ is the probability that action $a^t$ in state $s^t$ leads to state $s^{t+1}$ in the next time point;
  \item $R^t$ is the immediate reward received after receiving $a^t$ at $s^t$. It may also use information from $s^{t+1}$.
\end{itemize}

MDP is generally assumed for all methods under infinite time horizon. In an MDP, we often assume that a policy or treatment $\pi$ is used to make the decision. This is a distribution over actions given states $$\pi(s)={P}(A^t=a|S^t=s).$$ To balance the immediate reward and long-term reward, a discount factor $\gamma \in [0,1]$ is often used. Our goal is to find a policy $\pi$ that will maximize some cumulative version of rewards over time $t$:
$$\mathbb{E}\left[\sum^{\infty}_{k=0}\gamma^{k}R^{t+k}\right],$$
where we choose $a^t=\pi(s^t)$. Hence, when $\gamma$ is small, we only focus on the eventual reward when $t$ approaches infinity, while when $\gamma$ approaches 1, after proper normalization, we are interested in the long-term average reward \citep{meyn1997policy, liao2020batch}. Based on these definitions, we introduce the value function and the Q-function, which are fundamental concepts in a reinforcement learning task.

\textbf{Definition 3: (Value Function)} The state-value function $V_t^{\pi}(s)$ is the expected cumulative reward starting from state $s$, and then following policy $\pi$
$$V_t^{\pi}(s)=\mathbb{E}\left[\sum^{\infty}_{k=0}\gamma^{k}R^{t+k}|S^t=s\right].$$

\textbf{Definition 4: (Action Value Function)} The action-value function $Q^{\pi}_t(s,a)$ is the expected cumulative reward starting from state $s$, taking action $a$ and then following policy $\pi$
$$Q^{\pi}_t(s,a)=\mathbb{E}\left[\sum^{\infty}_{k=0}\gamma^{k}R^{t+k}|S^t=s,A^t=a\right].$$

Based on these definitions, we can treat both V-function and Q-function as estimates of how good a policy is for a patient in any given state. The only difference is whether we know the action at time $t$. In an MDP, $V_t^{\pi}(s)$ and $Q^{\pi}_t(s,a)$ do not depend on $t$ and are finite when $\gamma<1$ \citep{hinderer2005algorithms}. By solving a policy that maximizes these quantities, we essentially achieve the goal of personalized medicine. However, this is not trivial considering the dynamics over a long period, which can be difficult to model. Hence, the \emph{Bellman equation} becomes an important tool. The following are two versions of the Bellman equation corresponding to the value and Q-functions. Both of them use the property that they can be written in an integrative fashion in the MDP setting:
$$V^{\pi}(s)=\mathbb{E}[R^{t}+\gamma V^{\pi}(S^{t+1})|S^t=s],$$
$$Q^{\pi}(s,a)=\mathbb{E}[R^{t}+\gamma Q^{\pi}(S^{t+1},A^{t+1})|S^t=s,A^t=a].$$

The best treatment policy can be defined using either the value function, with $V^*(s)=\max_{\pi}V^{\pi}(s)$, or the Q-function with $Q^*(s,a)=\max_{\pi}Q^{\pi}(s,a)$. The two are connected via $V^*(s)=\max_{a}Q^*(s,a)$. Additionally, $Q^*(s,a)$ is unique and must satisfy the well-known \emph{Bellman optimality equation} \citep{puterman2014markov}:
\begin{equation}
Q^*(s,a)=\mathbb{E}_{S^{t+1}|s,a}\left[R^t+\gamma\max_{a'}Q^*(S^{t+1}, a')|S^t=s,A^t=a\right]. \label{bell}
\end{equation}
We refer to the policy $\pi^{*}$ that attains these maximums as the optimal policy, hence $Q^{\pi^{*}}(s, a) = Q^*(s,a)$. Equivalently, we can define the Bellman operator $\mathcal{B}$ of the value function as
\begin{equation}
    \mathcal{B}V(s)=\max_{a}\mathbb{E}_{S^{t+1}|s,a}\left[R^t+\gamma V(S^{t+1})|S^t=s,A^t=a\right]. \label{bellop}
\end{equation}
Then $V^{\pi^*}(s)$ is the fixed point of $\mathcal{B}$, and $\pi^{*}$ can be solved correspondingly. 

Before concluding this section, we want to remark that the decision maker's behavior in MDP during the data generating process is also crucial. Some conditions are needed to ensure that we can adequately estimate the treatment effects and make casual conclusions \citep{hernan2010causal}. However, due to the scope of this book chapter, we do not discuss causality issues in detail.
\begin{itemize}
 \item \textbf{Positivity:} ${P}(A^t=a^t|S^t=s^t)>0$ for all $a^t\in \mathcal{A}, s^t\in \mathcal{S}$, and all $t$. 
\item \textbf{Consistency:} Let $R(a^t)$ denote the potential outcome at time $t$ if receiving action $a^t$, then $R^t=R(a^t)$ if $a^t$ is actually used.
\item \textbf{Strong Ignorability:} $ \{R(a^t): a^t\in \mathcal{A} \} \perp \!\!\! \perp A^t \, | \, S^t$ for all $t$.
\end{itemize}

As a brief interpretation, we note that the positivity assumption ensures that all actions in the action space could be chosen at each time $t$ with a probability larger than 0. This allows the comparison among different actions. The consistency assumption guarantees that the estimates are valid, and the strong ignorability assumption suggests that there are no unmeasured confounders in the dataset. 

\section{Q-learning for Finite Horizon}\label{Qlearn}

Q-learning is one of the most important breakthroughs in reinforcement learning \citep{sutton2018reinforcement}. In the context of personalized medicine, it has been widely used in estimating the optimal treatment regime \citep{murphy2005generalization, robins2004optimal, zhao2009reinforcement, nahum2012q, kosorok2015adaptive} of finite-horizon problems where the number of decision stages is finite. Q-learning is also a popular approach in infinite horizon settings and achieves success in mobile health applications. Dynamic programming is a major technique to solve Q-learning problems without explicitly modeling the entire environment dynamics. In the following, we shall introduce Q-learning and its applications in the finite horizon setting. Then, we review some extensions of Q-learning in the infinite-horizon settings.

\subsection{Additional Notation}\label{Q_func}

In finite-horizon settings, data concerning personalized medicine are usually collected from a multi-stage clinical trial or longitudinal observational studies on the disease of interest \citep{clifton2020q}. When the number of stages is small, the Markov property is unnecessary. However, standard causal assumptions highlighted previously (positivity, consistency, and strong ignorability) are still assumed \citep{goldberg2012q} to make estimation and inference valid. An optimal dynamic treatment regime (DTR) aims to find a sequence of decision rules that assign treatments at each stage based on a patient's baseline characteristics and historical information. Some additional notations should be introduced to avoid confusion and allow higher-order dependencies during the decision process. 

Suppose we have pre-specified finite $T$ decision points, indexed by $t=1,2,...,T$. Note that in this setting, the time index starts with $1$ rather than 0, consistent with the literature. We denote $\bar A^t=(A^1,A^2,...,A^t)$ as a possible treatment trajectory administered through decision point $t$, and denote $\bar S^t=(S^1,S^2,...,S^t)$ as the historical information of the patient leading up to $t$. The realized treatment trajectory and historical information are denoted as $\bar a^t=(a^1,a^2,...,a^t)$ and $\bar s^t=(s^1,s^2,...,s^t)$, respectively. The immediate observed reward $R^t$ is slightly different here because the Markov property is not assumed. It considers all the previous history leading up to $t$. Another difference is that $\gamma=1$ under the finite horizon, as the cumulative reward will surely converge. Therefore, the goal is to find the optimal DTR to maximize the cumulative reward from $t=1$ to $T$, i.e., $\sum^T_{t=1}R^t$. 

A dynamic treatment regime, in this case, is defined as $\boldsymbol \pi=(\pi_1,...,\pi_T)$, which forms a sequence of policies to treat a patient over time. Here, the decision rules may depend on $t$, which is different from Markov settings. They can incorporate all information leading up to $t$. The optimal DTR $\boldsymbol {\pi^*}=(\pi_1^{*},\pi_2^{*},...,\pi_T^{*})$ is naturally defined as the DTR that satisfies, for all $\boldsymbol \pi$,
\begin{equation}
\mathbb{E}\left(\sum^T_{t=1}R^t_{\pi_t}\right)\leq \mathbb{E} \left(\sum^T_{t=1}R^t_{\pi_t^{*}} \right).
\end{equation}

\subsection{Backward Induction}

To estimate $\boldsymbol {\pi^{*}}$, Q-learning adopts a backward induction mechanism, moving from the final decision stage back to the first stage. We shall describe this process conceptually instead of giving details of the sample version since the finite-horizon setting is not the focus. Essentially, the algorithm is motivated by two facts. First, the best policy that maximizes the potential reward at the final stage can be learned by knowing the covariate at the $(T-1)$th stage. Secondly, the optimal treatment decision can be induced at any given time point by knowing the best decision at the next time point \citep{zhao2015new}. To better illustrate how the algorithm works, we define the following quantities. They are analogs of similar quantities introduced previously. At the final stage $T$, we have

\begin{align}
  Q_T(\bar s^T,\bar a^T) &=\mathbb{E}\left(\sum^T_{t=1}R^t|\bar S^T=\bar s^T,\bar A^T=\bar a^T\right),  \notag \\
\pi_T^{*}(\bar s^T, \bar a^{T-1})&=\text{argmax}_{a^T}Q_T(\bar s^T,\bar a^{T-1},a^T), \notag \\
\text{and} \quad V_T(\bar s^T,\bar a^{T-1}) &=\text{max}_{a^T}Q_T(\bar s^T,\bar a^{T-1},a^T)  \notag . 
\end{align}

For stages $t=T-1,...,1$, we perform the inductions:
\begin{align}
  Q_t(\bar s^t,\bar a^t) &=\mathbb{E} \big[ V_{t+1}(\bar s^t,S^{t+1},\bar a^t)|\bar S^t=\bar s^t,\bar A^t=\bar a^t \big] \label{q2}\\
\pi_t^{*}(\bar s^t, \bar a^{t-1})&=\text{argmax}_{a^t}Q_t(\bar s^t,\bar a^{t-1},a^t), \notag\\
\text{and} \quad V_t(\bar s^t,\bar a^{t-1}) &=\text{max}_{a^t}Q_t(\bar s^t,\bar a^{t-1},a^t). \notag
\end{align}

To briefly explain these notations, we start from the final stage $T$ and posit a model for $Q_T(\bar s^T, \bar a^{T})$, such as a linear model, SVM, or random forests \citep{zhao2009reinforcement}. We treat the observed cumulative reward $\sum^T_{t=1}R^t$ as the response, and $\bar s^T,\bar a^{T-1}$ as the covariates. After obtaining the model estimation using the observed data, we compare different treatment options to obtain $\hat \pi^{*}_{T}$ for a patient with history $\bar S^T=\bar s^T, \bar A^{T-1}=\bar a^{T-1}$ at decision point $T$. This allows us to estimate the best value function if only the last time-point requires a decision. Similar procedures can be adopted for the previous time-point $T-1$. Note that by Equation \eqref{q2}, the Q-function can be estimated with the already estimated value function in the next stage. Hence, we can compare and estimate the best treatment $\hat \pi^{*}_{T-1}$ at a previous time point. Recursively performing these operations, the estimated optimal treatment regime is then $\boldsymbol {\hat\pi^*}=(\hat\pi^{*}_{1},...,\hat\pi^{*}_{T})$.

In the finite horizon setting, Q-learning has gained tremendous popularity due to its simplicity of implementation and overall good performance \citep{kosorok2019precision}. However, it has limitations. For example, it is sensitive to model misspecifications. If the posited model for Q-functions is misspecified, the performance may suffer significantly. This is because the misspecification errors at each stage will be backpropagated and accumulated back to the first stage. Some alternatives have been proposed, such as Advantage Learning (A-learning, \cite{murphy2003optimal}), which estimates the optimal regime by modeling the difference in outcomes between two treatment options \citep{blatt2004learning}. Thus, it does not require the entire Q-function to be specified, making it more robust to model misspecification \citep{schulte2014q}. Other methods include the robust Q-learning \citep{ertefaie2021robust} that allows estimating nuisance parameters using data-adaptive techniques to address residual confounding and efficiency loss. 

The backward induction algorithm for finite-horizon Q-learning can be extended to infinite horizon settings in principle. This again requires accurate estimations of the Q-function. Hence, it is more prevalent in an environment where we can quickly and easily obtain new samples through experiments under a specific policy. Such a strategy may not be realistic for a practical situation in medical science. To deal with this issue, in Section \ref{V_GCQ}, we will review other Q-learning algorithms in infinite horizon settings, often referred to as the infinite-horizon Q-learning, first proposed by \cite{watkins1992q}. 

\section{Temporal Difference Learning} \label{sec:tdlearn}

There are three representative approaches for policy learning: dynamic programming, Monte-Carlo, and Temporal Difference (TD) learning \citep{sutton1988learning}. Among them, TD learning is a classical and powerful approach and often serves as the foundation of modern reinforcement learning. It has achieved great success in theoretical foundations and real applications \citep{dann2014policy}. Instead of modeling the entire dynamic for policy learning, TD learning is a model-free approach that can directly perform policy evaluation and optimization given any unknown transition Markov kernel. The advantage of a model-free approach is the high sample efficiency, which is highly desirable in policy learning. 

\subsection{On-Policy vs. Off-Policy Learning}

It would be helpful to clarify the difference between two notions, on-policy and off-policy since a specific TD learning approach depends on the setting. The difference can be roughly understood as if the target of our estimation is based on the observed treatment decision behavior or some other treatment strategy, e.g., the optimal one. For example, the Q-learning algorithms we introduced previously are off-policy learning. Their estimation target is the $Q$-function under the optimal treatment decision function $\pi$, which is rarely realized in an observational study. In contrast, another algorithm called State–Action–Reward–State–Action (SARSA) is interested in learning the $Q$-function if we follow the current observed behavior of treatment decision-making. Hence, it is evident that SARSA cannot be used to learn the optimal policy unless the patient is already implementing it in practice. Off-policy is likely the dominating case in medical-related studies since patients and doctors usually do not know the optimal policy in advance. However, in online gaming, the on-policy setting is quite frequent since one can manipulate machines to implement any new decision rule. This allows on-policy algorithms such as SARSA to update the policy and the corresponding $Q$-functions progressively. But in offline settings where data are already collected before any analysis, Q-learning can be more useful than SARSA. We shall discuss the online learning setting in Section \ref{sec:online}.

One difficulty involved in off-policy learning is that one must account for the change of distribution of states $S$ (distribution shift) when the target policy being estimated is no longer the one being used. This is especially true for offline settings since the observed (or implemented) policy can not be changed once data have been collected. Fundamentally, this is an observational study problem that often appears in statistics. One may expect some form of propensity score adjusting approaches to be implemented to handle the mismatch of the distributions. These ideas will be illustrated in detail in Section \ref{sec:TD_off}. In the following, we shall first focus on the on-policy setting to introduce the intuition behind TD learning and then discuss its off-policy generalization via importance sampling. 

\subsection{On-Policy TD Learning}

In this section, we introduce the on-policy TD learning, where the data is collected in an online fashion. In the on-policy setting, the agent can directly interact with the data-generating environment and continuously collect data from the environment online. This is distinguished from the off-policy setting, where the data is pre-collected, and the agent cannot interact with the environment. Typically, we observe a single sequence of streaming data consisting of the current state, next state, current action, and immediate reward at each time-point $t$.

The reinforcement learning problem can also be viewed as a traditional regression framework. The goal of TD learning is to minimize the discrepancy between the true value function $V^{\pi}$ and its function approximation counterpart $V^{\pi}_{\theta}$.
One of the most commonly used loss functions for measuring the discrepancy is the squared error (SE),
\begin{equation}
\operatorname{SE}_{t}(\theta)=\left[V_{\theta}^{\pi}\left(s^{t}\right)-V^{\pi}\left(s^{t}\right)\right]^{2}.
\label{MSE}
\end{equation}
Using a stochastic gradient descent with step size $\alpha$, the parameter update can be understood as, 
\begin{equation}
\begin{aligned}
\theta_{t+1} &=\theta_t -\alpha \nabla \operatorname{SE}_{t}(\theta) \\
&=\theta_t-\alpha\left[V_{\theta}^{\pi}\left(s^{t}\right)-V^{\pi}\left(s^{t}\right)\right] \nabla_{\theta_{t}} V_{\theta_{t}}\left(s^{t}\right).
\end{aligned}
\label{gradient_no_td}
\end{equation}
Still, Equation \eqref{gradient_no_td} can not be directly computed as the true value function $V^{\pi}$ is unknown. One natural solution is to collect Monte-Carlo samples to approximate $V^{\pi}$. In particular, one can repeat the experiments by implementing the policy $\pi$ and collecting a total of $m$ sample trajectories given state $s^t$. This is likely only possible in an online setting, as we could directly interact with the environment and collect the data. If we denote the empirical discounted sum of rewards from state $s^t$ as $\{V^{\pi}_{i}(s^t)\}^{m}_{i=1}$, then the equation \eqref{MSE} is feasible to be optimized and reformed to be

\begin{equation}
\operatorname{SE-MC}_{t}(\theta)=\frac{1}{m}\sum_{i=1}^{m}\left[V_{\theta}^{\pi}\left(s^t\right)-V^{\pi}_{i}(s^t)\right]^{2}.
\label{MSE2}
\end{equation}

However, \cite{sutton1988learning} indicates that $V^{\pi}_{i}(s^t)$ can be difficult to collect. To explicitly perform this, the experiments have to be repeated $m$ times for each state which is only valid in simulated environments but not in practice due to the cost and ethics concerns, especially for medical-related studies. Another main drawback of this Monte-Carlo-based approach is the curse of high variance. The variance of the samples $\{V^{\pi}_{i}(s^t)\}^{m}_{i=0}$ tends to be very large for a large number of decision time points. This is particularly severe in infinite-horizon settings. 

A stable and reliable alternative is needed. \cite{sutton1988learning} propose to view the true value function $V^{\pi}(s^{t})$ at the decision step $t$ as the composition of the immediate reward $R^t$ and the discounted value function approximation $\gamma V_{\theta}^{\pi}(s^{t+1})$ at step $t+1$. This composition is also called the value function one-step prediction, i.e.,
\begin{equation}
V^{\pi}\left(s^{t}\right) \approx R^{t}+\gamma V_{\theta}^{\pi}\left(s^{t+1}\right).
\label{onestep}
\end{equation}
The main idea of this update is motivated by the well-known Bellman equation. As the approximation of the true value function is obtained, one can proceed to minimize equation \eqref{MSE} with the approximation form equation \eqref{onestep}. The gradient descent can be calculated as follows, 
\begin{equation}
\begin{aligned}
\theta_{t+1} &=\theta_t -\alpha \nabla \operatorname{SE}_{t}(\theta) \\
&\approx \theta_t-\alpha\left[V_{\theta_{t}}^{\pi}\left(s^{t}\right)-R^{t}-\gamma V^{\pi}_{\theta_{t}}\left(s^{t+1}\right)\right] \nabla_{\theta_{t}} V^{\pi}_{\theta_{t}}\left(s^{t}\right),
\end{aligned}
\label{function_td}
\end{equation}
where $\theta_{t}$ is the updated parameter at step $t$ and $\alpha$ is the learning rate. Here the term  
$$
\delta^{t}=V^{\pi}_{\theta_{t}}\left(s^{t}\right) -\left[R^{t}+\gamma V^{\pi}_{\theta_{t}}\left(s^{t+1}\right)\right]
$$ 
is called temporal difference error, and $e_{t} \coloneqq \nabla_{\theta_{t}} V_{\theta_{t}}\left(s^{t}\right) = \phi\left(s^{t}\right)$ is the feature basis. At each updating step, the temporal difference $\delta^{t}$ indirectly quantifies the discrepancy between the true value function and its function approximation counterpart. Consequently, the equation \eqref{function_td} turns out to be
\begin{equation}
\begin{array}{l}
\theta_{t+1}=\theta_{t}-\alpha \delta^{t} e_t.
\end{array}
\label{linear_TD_eq}
\end{equation}
One main advantage of this strategy is to circumvent the curse of high variance within the Monte-Carlo sampling methods. From the gradient accumulation point of view, this step-wise gradient updating $\nabla_{\theta_{t}} V_{\theta_{t}}$ is actually propagating the gradient from the end decision step $t_{\text{end}}$ during the learning process to the staring decision step $t_{\text{start}}=0$. 

\subsection{Generalization to Off-Policy Learning}\label{sec:TD_off}

The TD-learning method discussed in the last section is usually referred to as on-policy TD-learning. The notion of the on-policy represents the reinforcement learning tasks that target estimating the value function $V^{\pi}$ in the scenario that the observed samples are collected when the policy $\pi$ is executed.

Another strand of the reinforcement learning task is off-policy learning, which is the dominating case in clinical trials and precision medicine. The observed samples collected are not following the policy $\pi$ but other exploration policies such as the randomization policy. Although the goal of estimating the value function $V^{\pi}$ remains the same as in on-policy learning, off-policy learning faces additional challenges, especially for distribution mismatching and the curse of high variance in infinite horizon settings \citep{jiang2016doubly}. 

\begin{figure}[h]
\label{reweight_plot}
\centering
\includegraphics[scale=0.6]{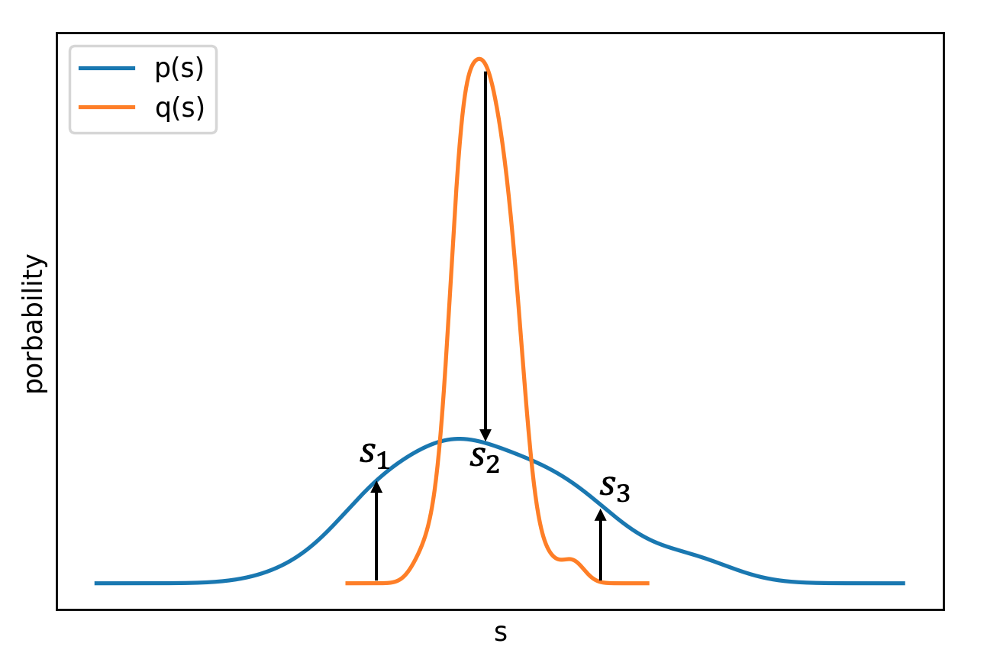}
\caption{Importance Sampling: Samples drawn from $q(x)$ are re-weighted by the importance weight $p(s_i)/q(s_i)$ for $i=1,2,3$ to behave like samples from $p(x)$.}
\end{figure}

The most important tool for off-policy learning is importance sampling. To further explain this notion, let us consider the problem of estimating $\mathbb E_{p}[f(S)]$ of a function $f \in \mathcal{F}$ with the input $S \sim p$, where $p$ is any distribution. If we can get access to the Monte-Carlo samples following the distribution $p$, the expectation $\mathbb E_{p}[f(S)]$ can be approximated by its mean estimator. However, this fails if the samples can not be collected with respect to the distribution $p$ but only for another distribution $q$. The importance sampling is a key tool for correcting the sample distribution $p$ to the distribution $q$. Denote the Monte-Carlo samples as $\{s_{i}\}^{m}_{i=1}$ following the distribution $q$, then the expectation $\mathbb{E}_{p}[f(S)]$ can be re-weighting according to the ``mismatch" of the density for two distributions, i.e., 
$$
\mathbb{E}_{p}[f(S)]=\mathbb{E}_{q}\left[\frac{p(S)}{q(S)} f(X)\right].
$$
Note that the right-hand side of the expectation is taken with respect to the distribution $q$. An immediate approximation of the above expectation is 
$$
\frac{1}{m} \sum_{i=1}^{m} \frac{p\left(s_{i}\right)}{q\left(s_{i}\right)} f\left(s_{i}\right).
$$
If the positive assumption holds, that is $q(s)$ is positive for any $s \in \mathcal{S}$, the ``mismatch'' of the two density $p(s_i)/q(s_i)$ is also known as the inverse probability weighting. We illustrate the inverse probability weighting in Figure \ref{reweight_plot}.

\textit{Challenges of Off-Policy Estimation} Off-policy estimation also raises many difficulties. In the on-policy setting, as the behavior policy distribution is equivalent to the target policy distribution, i.e., $\pi_{B} = \pi$, the state distribution $d$ involved under the target policy $d^{\pi}$ is the same as the state distribution involves under the behavior policy $d^{\pi_{B}}$, i.e., $d = d^{\pi} = d^{\pi_{B}}$. This implies that $d$ is following a stationary distribution. In contrast, in the off-policy setting that $\pi \neq \pi_{B}$, the state distributions under the target and behavior policy are different, i.e., $d^{\pi} \neq d^{\pi_{B}}$. Hence, the state distribution $d$ is not stationary in general. This makes the estimation quite unstable. Even worse, the estimated value function under off-policy learning could be divergent from the one in an on-policy learning \citep{dann2014policy}. A straightforward example is the off-policy mean squared projected Bellman error (MSPBE), 
\begin{equation}
\operatorname{MSPBE}({\theta}) \coloneqq \left\|{V}_{{\theta}}- \mathcal{B}^{\pi} {V}_{{\theta}}\right\|_{{D}_{\pi_{B}}}^{2}.
\end{equation}
If the state distribution $d^{\pi_{B}}$ is different from the state distribution $d^{\pi}$, the estimation results will exhibit significant discrepancies. In the following, we will review an generalization of the TD-learning to the off-policy setting via importance sampling.

\textit{Off-Policy TD Learning} Recall the parameter updates in an on-policy TD learning \eqref{linear_TD_eq}.
At decision step $t$, one needs to approximate the expectation $\mathbb{E}_{\pi, \mathcal{P}, d^{\pi}}[\delta_{t}\phi(s^{t})]$. The same updates can still be generalized to off-policy settings if the samples follow $\pi_B$. This motivates the following equations
\begin{equation}
\begin{aligned}
\mathbb{E}_{\pi, \mathcal{P}, d^{\pi}}\left[\delta_{t} \phi\left(s^{t}\right)\right] &= \mathbb{E}_{s^{t+1}|s^{t},a^{t}; \, s^{t} \sim d^{\pi}}  \left[ \pi \left(a^{t} \mid s^{t}\right) d^{\pi}\left(s^{t}\right) \delta_{t} \phi\left(s^{t}\right) \right] \\
&= \mathbb{E}_{s^{t+1}|s^{t},a^{t}; \,  s^{t} \sim d^{\pi_{B}}}  \left[ \pi \left(a^{t} \mid s^{t}\right) d^{\pi_{B}}\left(s^{t}\right) \delta_{t} \phi\left(s^{t}\right) \right] \\
&= \mathbb{E}_{s^{t+1}|s^{t},a^{t}; \,  s^{t} \sim d^{\pi_{B}}}  \left[ \pi_{B} \left(a^{t} \mid s^{t}\right) d^{\pi_{B}}\left(s^{t}\right) \frac{\pi\left(a^{t} \mid s^{t}\right)}{\pi_{B}\left(a^{t} \mid s^{t}\right)} \delta_{t} \phi\left(s^{t}\right) \right] \\
&=\mathbb{E}_{\pi_{B}, \mathcal{P}, d^{\pi_{B}}}\left[\frac{\pi\left(a^{t} \mid s^{t}\right)}{\pi_{B}\left(a^{t} \mid s^{t}\right)} \delta_{t} {\phi}\left(s^{t}\right)\right].
\end{aligned}
\end{equation}
Through the above procedure, the expectation taken with respect to the target policy $\pi$ is corrected to the expectation taken with respect to the behavior policy $\pi_{B}$. This is connected by the Radon–Nikodym derivative $\pi\left(a^{t} \mid s^{t}\right)/{\pi_{B}\left(a^{t} \mid s^{t}\right)} $. To execute the off-policy TD learning parameter updates, we have the following, which is the off-policy counterpart of the parameter updates in \eqref{linear_TD_eq},
\begin{align*}
\theta_{t+1}=\theta_{t}-\alpha \frac{\pi\left(a^{t} \mid s^{t}\right)}{\pi_{B}\left(a^{t} \mid s^{t}\right)}\delta_{t} e_{t} ,
\label{off_linear_TD}
\end{align*}
Note that we approximate the true value function $V^{\pi}$ by $V^{\pi}_{\theta}$, TD learning can diverge or be unstable when adopting a non-linear function approximation, especially in off-policy learning \citep{tsitsiklis1997analysis}. Hence, TD learning is also called a semi-gradient method. True gradient methods \citep{zhang2020deep}, such as residual gradient methods, have also been proposed and extended to off-policy settings to avoid this drawback. We will review one of the typical true gradient methods in Section \ref{resg}. Before that, we will review a few recent developments in the statistical literature that are closely connected with some of the properties we have introduced already. 

\section{Temporal Difference and Estimating Equations}\label{V_GCQ}

The temporal difference error and its associated properties are extensively used. Two recent methods \citep{ertefaie2018constructing, luckett2020estimating} utilize this tool and formulate the optimal treatment regime problem in an estimating equation fashion. Note that following the Bellman equation \eqref{bell}, the optimal Q-function satisfies
\begin{equation}
Q^*(s,a)=\mathbb{E}_{S^{t+1}|s,a}\left[R^t+\gamma\max_{ a^{\prime}}Q^*(S^{t+1}, a^{\prime})|S^t=s,A^t=a\right]. \notag
\label{bellman}
\end{equation}
The optimal policy is $\pi^*(s)=\argmax_{a}Q^*(s,a)$. Instead of iteratively updating the $Q$ function as we introduced earlier, \cite{ertefaie2018constructing} proposed a Greedy Gradient Q-learning (GGQ). The key insight is to utilize the temporal difference error defined as
$$\delta^{t+1}(\theta)=R^t+\gamma\max_{a'}Q^*(S^{t+1},a')-Q^*(s,a),$$
Suppose $Q(s,a)=\theta^T\varphi(s,a)$, where $\varphi(s,a)$ is a set of feature basis over state-action pair $(s,a)$, $\theta$ is a $p$-dimensional parameter vector of parameters. Let $\theta^\ast$ be the true parameter vector that attains the optimal value $Q^\ast(s,a)={\theta^\ast}^T\varphi(s,a)$. Then inspired by the property of temporal difference error that $\mathbb{E}[\delta^{t+1}(\theta^\ast)|S^t=s,A^t=a]=0$ for any $t\geq 0$ under the true $\theta^\ast$, the estimating equations of $\theta$ can be directly obtained as
\begin{equation}
    \mathbb{E}\left\{\sum^{T}_{t=0}\delta^{t+1}(\theta) \nabla_{\theta}Q_{\theta}(S^t, A^t) \right\}=0,
    \label{estimate_eq}
\end{equation}
where $\nabla_{\theta}Q_{\theta}(s,a)$ accounts for the influence of basis. Based on the linear assumption, $\nabla_{\theta}Q_{\theta}(s,a) = \varphi(s, a)$. Hence we can solve the sample version of the sum of trajectory loss 
$$\sum^n_{i=1}\sum^T_{t=0}\delta_{i}^{t+1}(\theta)\varphi(S^t_i,A^t_i)$$
to obtain an unbiased estimator $\hat\theta$ for $\theta^\ast$, and the corresponding estimated optimal policy is $\hat\pi^*(s)=\argmax_{a}Q_{\hat\theta}(s,a)$.

GGQ allows the construction of confidence intervals for the mean outcome difference between the optimal policy and any other policy. Another advantage is that the significance of variables can be tested in the optimal policy by checking whether the confidence interval for the corresponding parameter contains zero. However, the estimation equation \eqref{estimate_eq} contains an non-smooth $\text{max}$ operator, which makes the estimation difficult without large amounts of data \citep{laber2014dynamic}. Essentially, the Q-learning framework is a greedy method as it always tends to select the best arm for each decision stage. When the dynamic environment is complex, the greedy action often leads to sub-optimality \citep{dann2014policy}.

\cite{luckett2020estimating} developed another approach called the V-learning. Although it is fundamentally another application of the Bellman equation and the temporal difference error, it maximizes the value function rather than the action-value function. It can also be viewed as an estimating equation approach. Denote the propensity score at stage $t$ as $\mu^t(a^t,s^t)=P(A^t=a^t|S^t=s^t)$. In a micro-randomized trial, $\mu^t(a^t,s^t)$ is known. However, it must be estimated from the data for an observational study. Under assumptions specified in Section \ref{note}, \cite{luckett2020estimating} define the value function as
\begin{equation}
V^{\pi}(s)=\sum^{\infty}_{k=0}\mathbb{E}\bigg[\gamma^{k}R^{t+k}\bigg\{\prod_{v=0}^k\frac{\pi(A^{v+t},S^{v+t})}{\mu^{v+t}(A^{v+t},S^{v+t})}\bigg\}\bigg\rvert S^t=s\bigg].
\end{equation}
For any $t$, the following importance-weighting variant of the Bellman optimality equation holds, 
\begin{equation}
    \mathbb{E}\bigg[\frac{\pi(A^t,S^t)}{\mu^t(A^t,S^t)}\big\{R^t+\gamma V^{\pi}(S^{t+1})-V^{\pi}(S^{t}) \big\}\phi(S^t)    \bigg]=0,
\end{equation}
where $\phi$ denotes the weight function of any state. Similar to GGQ, we can parameterize the state-value function $V^{\pi}(S_i)$ with $\theta$. Then the estimating equation is given by
\begin{equation}
    \Lambda_n(\pi,\theta)=\frac{1}{n}\sum^n_{i=1}\sum^{T}_{t=0}\frac{\pi(A_i^t,S_i^t)}{\mu^t(A_i^t,S_i^t)}
    \big\{R_i^t+\gamma V^{\pi}_{\theta}(S^{t+1}_i)-V^{\pi}_{\theta}(S^{t}_i)\big\}\nabla_{\theta}V^{\pi}_{\theta}(S^{t}_i),
    \label{offline-V}
\end{equation}
where $\nabla_{\theta}V^{\pi}_{\theta}(S_i)$ is the gradient of $V^{\pi}_{\theta}(S_i)$.

V-learning avoids the non-smooth max operator, which makes the optimization more stable. Compared with GGQ, V-learning considers a stochastic policy distribution, which can be more robust to unexpected situations \citep{zhou2021estimating}. Although both GGQ and V-learning are pioneer works in infinite horizon DTR estimations, many open problems remain, such as finding the most appropriate features for the parametric model, addressing the computational difficulties for big data, and quantifying uncertainty for policy evaluations.

\section{Residual Gradient Algorithm} \label{resg}

TD learning has achieved great success in both theoretical developments and real-world applications \citep{dann2014policy}. As we can see from Section \ref{sec:tdlearn}-\ref{V_GCQ}, the temporal difference error can be used in various ways. It usually requires approximations of the true value function. \cite{tsitsiklis1997analysis} show that TD learning approaches guarantee the convergence to the optimal policy with discrete state space and linear function approximation \citep{tsitsiklis1997analysis}. Unfortunately, TD learning can often be divergent in the case of non-linear function approximation, such as deep neural networks. In fact, there is no explicit objective function minimized for parameter updates in TD-learning.

\cite{baird1995residual} proposed a true gradient approach, aiming to minimize an explicit objective function for the true value function estimation under the function approximations. The goal is to estimate the optimal value function $V$ by the function approximator $V_{\theta}$. Define the Bellman operator as 
$$
\mathcal{B} V(s):=\max _{a} \mathbb{E}_{s^{\prime} \sim P(\cdot \mid s, a)}\left[R+\gamma V\left(s^{\prime}\right) \mid s, a\right],
$$
where $s^{\prime}$ is the next state for the current state $s$. Then \cite{baird1995residual} proposed to minimize the mean square Bellman residual,
$$
\text{MSBE} = \mathbb{E}_{s \sim \mu}\left[\left\{\mathcal{B} V(s)-V(s)\right\}^{2}\right],
$$
where $\mu$ is an observed state distribution. With function approximations and substituting the unknown transition kernel by its one-sample counterpart, the empirical MSBE with $m$ observed samples is 
\begin{align}
\widehat{\text{MSBE}}(\theta)\ &= \frac{1}{m} \sum_{i}\left[\widehat{\mathcal{B}}V_{\theta}(s_i) - V_{\theta}(s_{i})\right]^{2} \\
& = \frac{1}{m} \sum_{i}\left[\left\{ R_{i} + \gamma V_{\theta}\left(s^{\prime}_i \right)\right\}-V_{\theta}(s_{i})\right]^{2},
\end{align}
where $s^{\prime}$ is the next state of the state $s$. The gradient descent can be formulated as 
\begin{equation}
\begin{aligned}
\theta_{t+1} &=\theta_t -\alpha \nabla \widehat{\text{MSBE}}(\theta) \\
&=\theta_t-\alpha\left[R_i + \gamma V_{\theta_{t}}\left(s^{\prime}\right)-V_{\theta_{t}}\left(s\right)\right] \left[\gamma\nabla_{\theta_{t}} V_{\theta_{t}}\left(s^{\prime}\right) - \nabla_{\theta_{t}} V_{\theta_{t}}\left(s \right) \right],
\end{aligned}
\label{function_no_td}
\end{equation}
where $t$ is the iterative step different from the decision stage step in TD learning. According to the Bellman optimality equation, the MSBE will attain zero if the value function is optimal. Therefore, performing gradient descent on MSBE guarantees that MSBE will eventually converge to a local minimum. 

However, in a stochastic environment, the residual gradient algorithm encounters \emph{double sampling issues}. The double sampling issue leads to an empirical objective function biased from its population version. 
\begin{align}
\mathbb{E}_{s\sim d^{\pi_{B}}}\left[\widehat{\text{MSBE}}(\theta)\right]=\text{MSBE}(\theta) +\mathbb{E}_{s \sim d^{\pi_{B}}}\left[\operatorname{var}(\widehat{\mathcal{B}} V(s) \mid s)\right] \neq \text{MSBE}(\theta).
\end{align}
We will discuss more details of this double sample issue in section \ref{our_method}. As a result, the residual gradient is a biased algorithm but can be guaranteed to converge with non-linear function approximation in a stochastic environment. To correct the bias, \cite{baird1995residual} propose to sample the next state twice or more times given any current state. However, this is not practical in the real world. The next section will review a novel approach for resolving the double sampling issue using a functional space embedding framework. 

\begin{rmk}
For the algorithm implementation, the value function $V$ is usually linearly approximated as a weighted sum of a set of given basis functions $\phi_1,...,\phi_p$:
$
{V}_{\theta}(s)=\sum_{i=1}^{p} \theta_{i} \phi_{i}(s)
$.
The next step is to learn the set of weights $\theta = [\theta_1,...,\theta_p]$
corresponding to an approximation of the optimal value function $V$, and the value function is linear in $\theta$. The linear function approximations are widely applicable in reinforcement learning tasks as they are suitable for gradient descent and possess a quadratic error surface with a sole minimum. The flexibility is preserved by the choices of the basis function, including linear, polynomial, and Gaussian basis, to represent a very complex functional space. In addition to linear approximations, the value function can also be approximated using deep neural networks and random forests to gain more representation power. 
\end{rmk}

\section{Proximal Temporal Consistency Learning}\label{our_method}

As we have discussed in the last section, one of the biggest challenges for true gradient descent reinforcement learning is the double sampling issue \citep{baird1995residual}. To solve this issue, \cite{zhou2021estimating} leveraged the property of functional space embedding to resolve the double sampling issue very efficiently. 

In addition to the double sampling issue, some limitations exist for the previously reviewed methods for the applications to solve real-world problems in certain situations. First, Q-learning is a greedy algorithm and thus yields sub-optimal solutions. Meanwhile, policies induced by Q-learning are deterministic, which means they can only provide a single recommendation action at each decision point. This gives little flexibility to select treatments if that option is unavailable due to a temporary medication shortage or not available for other reasons. A possible remedy is to model the policy distribution stochastically. However, existing works typically model the policy distribution by Boltzmann distribution, which can quickly decay to a uniform distribution when the number of treatment options is too large \citep{min2014interpretable,zhou2021estimating}. This is a common issue in all existing approaches. 

We will discuss and address these two issues with the newly proposed Proximal Temporal Consistency Learning (pT-Learning) framework proposed by \cite{zhou2021estimating}. 

\subsection{Temporal Consistency and Functional Space Embedding }\label{sec:tempcon}

To address the double sampling issue, we need to correct the bias from the one-sample approximation with unknown Markov transitions. Following the Bellman optimality equation, if $\pi^*$ is the optimal policy, and $V^*$ is the corresponding optimal value function, then $(V^*, \pi^*)$ is a solution of the following temporal consistency equation for any $s$ and $a$:
\begin{equation}
    \mathbb{E}_{S^{t+1}}\big[R^t+\gamma V(S^{t+1})\big]-V(s)=0, \label{tempcon_org}
\end{equation}

The intuitive idea is to minimize the square loss of the temporal consistency error, which is the residual gradient algorithm. However, per our discussion, replacing the unknown transition kernel with its sample counterparts will lead to the double sampling issue. Instead, we can embed a Lebesgue measurable function class to the averaged temporal consistency error. Specifically, let's define a critic function $h(s,a)$ and formulate a novel embedding loss $\mathcal{L}_{\text{weight}}$ as follows,
\begin{equation}
 \mathcal{L}_{\text{weight}}(V,\pi,h):=\mathbb{E}_{S^t,A^t}\Big[h(S^t,A^t)\big\{ R^t+\gamma V(S^{t+1})-V(S^t) \big\}\Big].
\end{equation}

To better understand this embedding, the critic function introduced here is actually to fit the discrepancy of the left-hand side of the equation \eqref{tempcon_org}, and simultaneously promotes those sampled transition pairs with large temporal consistency error. Intuitively, the embedding loss circumvents the double sampling issue since it does not involve the additional variance (bias term) from one-sample bootstrapping. Fundamentally, the embedding loss $\mathcal{L}_{\text{weight}}$ offers compensation for insufficient sampling on $P(s'|s,a)$. The marginal distribution of the state-action pairs $d^{\pi}(s,a)$ and the transition kernel $P(s'|s,a)$ can be aggregated to joint distribution $p_{\pi}(s',s,a)$ by the linearity of expectations. Therefore, the embedding loss can be approximated using the sample-path transition pairs drawn from the joint distribution instead of directly approximating the unknown transition kernel. This explains why the embedding loss can resolve the double sampling issue.

Besides these facts, the critic function has nice properties in solving the optimization problem. First, $\mathcal{L}_{\text{weight}}(V, \pi)$ is a valid loss which can identify the optimal value function and optimal policy $(V^*,\pi^*)$ in that $\mathcal{L}_{\text{weight}}(V^*,\pi^*,h)=0$. Furthermore, if there exists $(\tilde V,\tilde \pi)$ such that $\mathcal{L}_{\text{weight}}(\tilde V,\tilde \pi,h)=0$, then $(\tilde V,\tilde \pi)$ satisfies the equation \eqref{tempcon_org}. Noticing that $\mathcal{L}_{\text{weight}}(V^*,\pi^*,h)=0$ holds for any $h$ in a bounded $L^2$ space, such property immediately transforms the original problem to the following minimax optimization problem
\begin{equation}
    \min_{V,\pi} \max_{h} \mathcal{L}_{\text{weight}}^2(V,\pi,h). \label{minimax}
\end{equation}

The minimax optimization of \eqref{minimax} gives a clear direction for estimating $(V^*,\pi^*)$. However, it is still not tractable, as the critic function could be arbitrary, making it infeasible to find a proper representation of $h(\cdot)$. However, if the observed reward $R^t$ is a function continuous over $(s^t, a^t)$, then the critic function $h^*(s,a)$ that maximize $\mathcal{L}_{\text{weight}}^2(V,\pi,h)$ is also continuous. Therefore, if we assume a continuous reward function, we can represent $h^*(s, a)$ in a bounded reproducing kernel Hilbert space (RKHS), and the minimax problem of \eqref{minimax} can be decoupled into the following single-stage minimization problem 
\begin{align}
     \min_{V,\pi}  \mathcal{L}_U=
     \mathbb{E}_{S^t,\tilde S^t,A^t,\tilde A^t,S^{t+1},\tilde S^{t+1}}\Big[\zeta \big\{\tilde{\mathcal{T}}_{\pi}(S^t,A^t,V)-V(S^t)\big\} K({S^t,A^t},{\tilde S^t, \tilde A^t})\big\{\tilde{\mathcal{T}}_{\pi}(S^t,A^t,V)-V(\tilde S^t)\big\} \Big],
     \label{mini}
\end{align}
where $\tilde{\mathcal{T}}_{\pi}(S^t,A^t,V)=R^t+\gamma V(S^{t+1})-\psi(\pi(A^t|S^t))$, $(\tilde S^t,\tilde A^t, \tilde S^{t+1})$ is an independent copy of the transition pair $(S^t,A^t,S^{t+1})$, and $K(\cdot,\cdot)$ is the universal kernel reproducing $h^*$. This property was explored by \cite{zhou2021estimating}.

Given the observed data $\mathcal{D}_n$ with a length $T$ trajectory, the above problem can be solved using a trajectory-based order-2 U-statistic estimator. Consequently, the total loss $\mathcal{L}_U$ can be aggregated as the empirical mean of $n$ $i.i.d.$ within-trajectory loss. The proposed sample estimator $\tilde {\mathcal{L}}_U$ is consistent, and its gradient can be approximated by the sampled transitions and optimized using any gradient descent algorithm. Thus, the proposed method achieves much flexibility regarding function approximations of $(V^*,\pi^*)$, allowing both linear and nonlinear approximations without the risk of divergence from the optimal solution. 

\subsection{A Proximal Bellman Operator with Sparse Policies} \label{sec:proxbell}

We have shown how the pT-Learning method resolves the double sampling issue. Equipped with this new tool, we turn to solve another practical issue when the number of treatment options is large. This involves constructing a proximal Bellman operator to facilitate smooth optimization and induce a sparse policy distribution. We first revisit the Bellman optimality equation in \eqref{bellop} from a policy explicit view. Suppose the policy $\pi$ follows a stochastic distribution, then the Bellman optimality equation can be rewritten as
\begin{equation}
    \mathcal{B}V(s):=\max_{\pi}\mathbb{E}_{a\sim \pi(.|s),S^{t+1}|s,a}[R^t+\gamma V(S^{t+1})|S^t=s,A^t=a]=V(s). \label{bellpi}
\end{equation}
A natural idea is to jointly optimize $V$ and $\pi$ for minimizing the difference between $\mathcal{B}V(s)$ and $V(s)$ for any $s$. However, this equation is nonlinear and contains a non-smooth $max$ operator again. This makes the estimation very difficult without a large number of samples. A proximal Bellman operator could circumvent this difficulty. Specifically, the operator adds a strongly convex and continuous component $d(\pi)$ to the dual of \eqref{bellpi}:
\begin{align}
 \mathcal{B}_{\lambda}V_{\lambda}(s) &=\max_{\pi}\sum_{a}[\langle \mathbb{E}_{S^{t+1}\sim p(.|s,a)}[R^t+\gamma V_{\lambda}(S^{t+1})],\pi(a|s)\rangle-\mu(\pi(a|s))+\lambda d(\pi(a|s))]\notag\\
 &= \max_{\pi}\mathbb{E}_{a\sim \pi(.|s)}[Q_{\lambda}(s,a)+\lambda \phi(\pi(a|s))], 
\label{appbell}
\end{align}
where $\mu(\cdot)$ is a constant function always equal to 0 in our case, $\phi(x)=d(x)/x$ with the proximity function $d(\cdot)$ and $Q_{\lambda}(s,a)=\mathbb{E}_{S^{t+1}|s,a}[R^t+\gamma V_{\lambda}(S^{t+1})]$. In particular, \cite{zhou2021estimating} consider a special case $d(x) = x\phi(x) = -\frac{x}{2}\log_{\kappa=0}(x)$ for achieving both smoothness and sparsity properties, where $
\log _{\kappa}(x)= \frac{1}{1-\kappa}\left(x^{1-\kappa}-1\right)
$.
\begin{rmk}
In general, the proximity function satisfying the following properties is valid for smoothness purposes. 1. Monotonicity: $\phi(x)$ is non-increasing; 2. Non-negativity: $\phi(1)=0$; 3. Differentiability: $\phi(x)$ is differentiable; 4. Expected Concavity: $x\phi(x)=d(x)$ is strictly concave. Many choices of $d(x)$ satisfy the above conditions. For example, when $d(x) = x\phi(x) = -x\log x$, the well-known Shannon entropy is recovered. 
\end{rmk}

There are several properties that we should highlight. First, the proximal Bellman operator \eqref{appbell} is a valid approximation for the Bellman operator $\mathcal{B}$, where the approximation bias can be bounded by
\begin{equation}
    \mathcal{B}_{\lambda}V(s)-\mathcal{B}V(s)=\frac{\lambda}{2}-\frac{\lambda}{2|\mathcal{K}(s)|},
\end{equation}
where $|\cdot|$ denotes the cardinality of a set, and $\mathcal{K}(s)$ represents the action set at state $s$, that is $\mathcal{K}(s)=\{a_{(i)}\in \mathcal{A}: Q_{\lambda}(s,a_{(i)})>\frac{1}{i}\sum^1_{j=1}Q_{\lambda}(s,a_{(j)})-\frac{\lambda}{i}\}$. Here $a_{(i)}$ is the action with the $i$-th largest state-action value, and $|\mathcal{K}(s)|\leq |\mathcal{A}|$ always holds for any $s$.

Additionally, $\mathcal{B}_{\lambda}$ is a smooth approximation for $\mathcal{B}V(s)$, which is a direct result of applying the Lagrange multiplier. Specifically, $\mathcal{B}_{\lambda}$ has a closed-form equivalence
\begin{equation}
 \mathcal{B}_{\lambda}V_{\lambda}(s)=\frac{\lambda}{2}\left\{ 1-\sum_{a\in \mathcal{K}(s)}\left[\left(\frac{\sum_{a\in \mathcal{K}(s)} Q_{\lambda}(s,a)}{\lambda |\mathcal{K}(s)|}-\frac{1}{|\mathcal{K}(s)|}\right)^2 -\left(\frac{Q_{\lambda}(s,a)}{\lambda}\right)^2\right] \right\}
\end{equation}
By transforming the Bellman operator into a smooth operator, great convenience could be gained when solving the optimization problem later. 

Besides gaining advantages in solving the optimization problem, the unique advantage of introducing $\mathcal{B}_{\lambda}$ lies in that it induces a sparse optimal policy distribution $\pi_{\lambda}(a|s)$, whose support set is a sparse subset of action space. Specifically, by applying $\mathcal{B}_{\lambda}$, we could gain a stochastic policy that only includes optimal and near-optimal policies while removing those options far from optimal. This can be illustrated by presenting $\pi_{\lambda}(a|s)$ in terms of the state-action value function $Q_{\lambda}(s,a)$ analytically, that is
\begin{equation}
    \pi_{\lambda}(a|s)=\bigg(\frac{Q_{\lambda}(s,a)}{\lambda}-\frac{\sum_{a\in \mathcal{K}(s)} Q_{\lambda}(s,a)}{\lambda\mathcal{K}(s)}+\frac{1}{|\mathcal{K}(s)|}\bigg)^+. \label{pmfpi}
\end{equation}
Notice that \eqref{pmfpi} gives a well-defined probability mass function in that $\sum_a \pi_{\lambda}(a|s)=1$. Also, the policy $\pi_{\lambda}(a|s)$ tends to assign a large probability to the action according to the rank of the state-action values. Moreover, sparsity can be shown by analyzing the support set of $\pi_{\lambda}(\cdot|s)$, which satisfies
\begin{equation}
    \Big\{a_{(i)}\in \mathcal{A}: Q_{\lambda}(s,a_{(i)})>\frac{1}{i}\sum^i_{j=1}Q_{\lambda}(s,a_{(j)})-\frac{\lambda}{i}\Big\}. \label{suppi}
\end{equation}
The inequality \eqref{suppi} indicates the sparsity parameter $\lambda$ controls the margins between the smallest action value and the others included in the support set. To better demonstrate how the mechanism works, we consider a toy example where there are three available actions at a decision point. When $\lambda$ is sufficiently small, such that $\lambda<Q_{\lambda}(s,a_{(1)})-Q_{\lambda}(s,a_{(2)})$, only $a_{(1)}$ will be contained in the support set. If $\lambda$ increases and falls into the range $[Q_{\lambda}(s,a_{(1)})-Q_{\lambda}(s,a_{(2)}),\sum^2_{i=1}Q_{\lambda}(s,a_{(i)})-2Q_{\lambda}(s,a_{(3)})]$, the support set would contain both $a_{(1)}$ and $a_{(2)}$ but eliminates action $a_{(3)}$. 

Therefore, the cardinality of the support set increases as $\lambda$ increases. This allows us to tune between the deterministic and stochastic policy models by selecting an appropriate $\lambda$ value. Furthermore, only treatments with rewards close to the optimal treatment will be included in the support set. Besides, the sparse property may also benefit healthcare providers and hospitals. First, when there are many treatment options, it may not be convenient to store all medications and resources simultaneously for the hospital. Only storing the most commonly used optimal and near-optimal treatments would be enough, allowing better medical management. Secondly, two or more medications may have similar effectiveness in some situations. As other methods can hardly show such properties, treatment plans with sparse property could give healthcare providers more flexibility in choosing the most appropriate treatment for the patient. 

Plugging in the proximal Bellman Operator $\mathcal{B}_{\lambda}$ and the induced $V_{\lambda}$ to the framework we discussed in Section \ref{sec:tempcon}, we can derive an proximal temporal consistency equation For any $s,a$, and $\lambda \in (0, \infty)$, if $\pi_{\lambda}^*$ is the optimal policy, and $V_{\lambda}^*$ is the corresponding optimal value function, then $(V_{\lambda}^*, \pi_{\lambda}^*)$ is a solution of the following proximal temporal consistency equation:
\begin{equation}
    \mathbb{E}_{S^{t+1}|s,a}\big[R^t+\gamma V_{\lambda}(S^{t+1})\big]-\lambda \phi(\pi_{\lambda}(a|s))-\Psi (s)+\psi(a|s)-V_{\lambda}(s)=0, \label{tempcon}
\end{equation}
where $\Psi(s)\in [-\lambda/2,0]$ and $\psi(a|s)\in [0,\infty)$ are Lagrangian multipliers with $\psi(a|s)\pi_{\lambda}(a|s)=0$. We call the discrepancy on the left-hand side pT-error. 

Note that the equation \eqref{tempcon} holds for any state-action pair $(s, a)$, the policy $\pi_{\lambda}$ can be estimated directly over the observed transition pairs. This local consistency property makes it easy to incorporate off-policy data. By leveraging this proximal temporal consistency property and combine with the functional embedding framework in Section \ref{sec:tempcon}, \cite{zhou2021estimating} proposed to minimize the following objective function to estimate the optimal policy and value function, that is
\begin{equation}
    \begin{split}
     \min_{V_{\lambda},\pi_{\lambda},\Psi,\psi}  \mathcal{L}_U=
     &\mathbb{E}_{S^t,\tilde S^t,A^t,\tilde A^t,S^{t+1},\tilde s^{t+1}}\big[(\tilde{\mathcal{T}}_{\pi_{\lambda}}(S^t,A^t,V_{\lambda})-\Psi(S^t)+\psi(A^t|S^t)-V_{\lambda}(S^t))\\
     & \cdot \zeta K({S^t,A^t},{\tilde S^t, \tilde A^t})(\tilde{\mathcal{T}}_{\pi_{\lambda}}(S^t,A^t,V_{\lambda})-\Psi(\tilde S^t)+\psi(\tilde A^t|\tilde S^t)-V_{\lambda}(\tilde S^t)) \big],\\
     s.t.\quad & \pi_{\lambda}(a|s)\cdot\psi(a|s)=0 \quad \text{and}\quad -\lambda/2\leq \Psi(s)\leq 0 \; \text{for all} \; s\in \mathcal{S}, a \in \mathcal{A}. \label{mini_U}
    \end{split} 
\end{equation}

For the implementation, \cite{zhou2021estimating} reformulate the minimization problem \eqref{mini} to be a nonlinear programming problem. The objective function is converted to a quadratic term with a nonlinear equality constraint and a nonlinear inequality constraint. Some computational burden is also alleviated by imposing certain restrictions so that the two Lagrangian functions $\Psi,\psi$ satisfy the constraints automatically. The corresponding algorithm is implemented in the R package \texttt{proximalDTR}.

\subsection{OhioT1DM Data Analysis}

\subsubsection{A mobile health study: OhioT1DM}\label{mdata}

We finally apply the pT-Learning method \citep{zhou2021estimating} to the OhioT1DM\footnote{The mHealth dataset OhioT1DM can be requested and downloaded in \url{https://ohio.qualtrics.com/
jfe/form/SV_02QtWEVm7ARIKIl.}} dataset presented in Section \ref{motex}. The dataset contains two groups of individuals with Type-1 diabetes, while each group contains six individuals aged 20 to 60. All patients were on insulin pump therapy with continuous glucose monitoring (CGM) sensors, and the life-event data were collected via a mobile phone app. Specifically, the dataset includes CGM blood glucose levels measured per 5 minutes, insulin dose level delivered to the patient, meal intakes, and corresponding carbohydrate estimates. Other features, such as self-reported times of work, sleep, and stress, are also available for all patients.

However, the physiological data of the first cohort were collected by \textit{Basis} sensor bands, while the second cohort used \textit{Empatica} sensor band. As a result, the heart rate measured per 5 minutes is available only for patients in the first cohort. The magnitude of acceleration aggregated per minute is only available for patients in the second cohort.

Based on the preliminary investigation, each patient has distinct blood glucose dynamics. Therefore, we follow the pre-processing strategy used in \cite{zhu2020causal} and treat the data of each patient as a single dataset. Additionally, we assume that we treat each day as an independent trajectory for each patient and estimate the optimal policy based on these trajectories. Therefore, we fully consider the difference between individuals and develop a personalized treatment plan for each patient.

We summarize the measurements over 30-min intervals such that the length of each trajectory is $T=48$. After removing missing samples and outliers, each dataset contains $n=15$ trajectories on average. For both cohorts, we study the individualized dose-finding problem by discretizing the continuous dose level into 14 disjoint intervals for intervention options, i.e., $A_i^t\in \{0=A_{(1)}<...<A_{(14)}=\max(A)\}$, and the grid is determined by the data corresponding to each patient. For the patient in first cohort, patient's states at each stage include the average blood glucose levels $S_{i,1}^t$, the average heart rate $S_{i,2}^t$ and the total carbohydrates intake $S_{i,3}^t$ from time $t-1$ to $t$. For patients in the second cohort, the states are the same as in the first cohort, except that the average magnitude of acceleration replaces the average heart rate.

The reward for this problem is a measure of the health status of the patient's glucose level, and we define the immediate reward as the average of the index of glycemic control,
$$R_i^t=-\frac{\mathbbm{1}(S_{i,1}^t>140)|S^t_{i,1}-140|^{1.1}+\mathbbm{1}(S_{i,1}^t<80)(S_{i,1}^t-80)^2}{30},
$$
where $R_i^t$ is non-positive, and a larger value is preferred. For this problem, when we summarize the reward over 30-min intervals, we would use the average reward during that 30-min period. Our goal is to maximize the discounted sum of reward $\mathbb{E}\sum_{t\geq 1}\gamma^{t-1}R^t$.

\subsubsection{ProximalDTR Package and Analysis Results}

We have specified the states, action space, and reward for this example. We use  \texttt{proximalDTR}\footnote{The \texttt{proximalDTR}  package is in \url{https://github.com/vincentskywalkers/ProximalDTR-}.} package for implements the pT-Learning method discussed in this section\footnote{The implementation code for analyses in this chapter is in \url{https://github.com/liyuhan529/Policy-Learning-for-Individulized-Treatment-Recommendation}.}. Applying the algorithm to the dataset, we solve for the optimal treatment policy. The algorithm generates the recommended treatment policy that enjoys the sparse property when given a new state. We show the usage of the package for one patient from the dataset (1st patient in the first cohort) in this section. 

After processing the data as described in subsection \ref{mdata}, we obtain a matrix $X$ with the state at each time point and two vectors showing the action chosen $A$ and observed reward $R$ at each time point. For this patient, if $A^t=0$, no insulin injection is taken at time $t$. If $A^t=i$, then the dose unit for insulin injection is in the interval $[9(i-1)/14, 9i/14]$. The format of the data is shown below.

\begin{lstlisting}[language=R]
> X[1:7,]
      glucose    heart meal
49  343.00000 65.50000    0
97  122.16667 60.33333    0
145 156.16667 71.66667    0
193 187.66667 71.16667    0
241  84.00000 74.66667    0
385  88.66667 65.16667    0
433  92.66667 67.00000    0
> A[1:7]
[1] 0 0 0 0 0 0 0
> R[1:7]
[1] -11.5114   0.0000  -0.7128  -2.3391  0.0000  -1.4444  0.0000
\end{lstlisting}

We make several remarks. First, the longitudinal state matrix $X$ should follow a certain format to be used for the package. Specifically, suppose there are $n$ trajectories in the dataset, the first $n$ rows of the matrix should be the first state, i.e., $S_i^0$ $(i=1,2,...,n)$ for the $n$ trajectories, the $(n+1)$th to $2n$th row should be the second state for each trajectory, etc. Secondly, $X$ has one more stage than $A$ and $R$ since at the end of the stage, we only observe $X$ but not $A$ or $R$. Finally, for this problem, the glucose and acceleration level in matrix $X$ and reward $R$ is the average value during the 30-min when transforming from the original dataset containing a measurement every 5 minutes.

After transforming the data into the required form, we use the \texttt{proximalDTR} function to fit the data:

\begin{lstlisting}[language=R]
> fit1 = proximalDTR(X = X, A=A, R=R, n_ID = 33, stage=48, gamma=0.9,
+       lambda.set = c(2),step.beta = 0.001, step.omega = 0.001,
+       desc.rate = 0.001, max.iter = 3000, max.iter.cv = 1000, bw= 1,
+       cv=F, trace =TRUE)
\end{lstlisting}

Here, \texttt{n\_ID} is the number of trajectories, \texttt{stage} is the length of each trajectory, \texttt{gamma} is the discount factor, and \texttt{lambda.set} is the candidate values of $\lambda$ that we will choose from by cross-validation. We can also specify a certain $\lambda$ value to obtain an appropriate sparsity level for the induced policy. Other parameters in the function are used to control convergence.

When given a new state for the patient, the algorithm will induce a treatment policy that enjoys sparse property by \texttt{predict\_rl} function. An example is shown below: if the patient is in the state where the glucose level is 200 mg/dL, the heart rate is 90 bpm, and the food intake is 0, the recommended best treatment for that patient is the 6th treatment (insulin injection with dose 3.21-3.86 units). The corresponding probability is to choose the 6th treatment with a probability of 0.8110 and the 1st treatment (insulin injection with a dose level less than 0.64 unit) with a probability of 0.1890. 

\begin{lstlisting}[language=R]
> pred.fit2 <- predict_rl(fit1, c(200,90,0))
> pred.fit2$prob
 [1] 0.0000 0.1890 0.0000 0.0000 0.0000 0.0000 0.8110 0.0000 0.0000 0.0000 
[11] 0.0000 0.0000 0.0000 0.0000
> pred.fit2$recommend.trt
[1] 6
\end{lstlisting}

\begin{figure}[htp]
\centering
\includegraphics[scale=0.7]{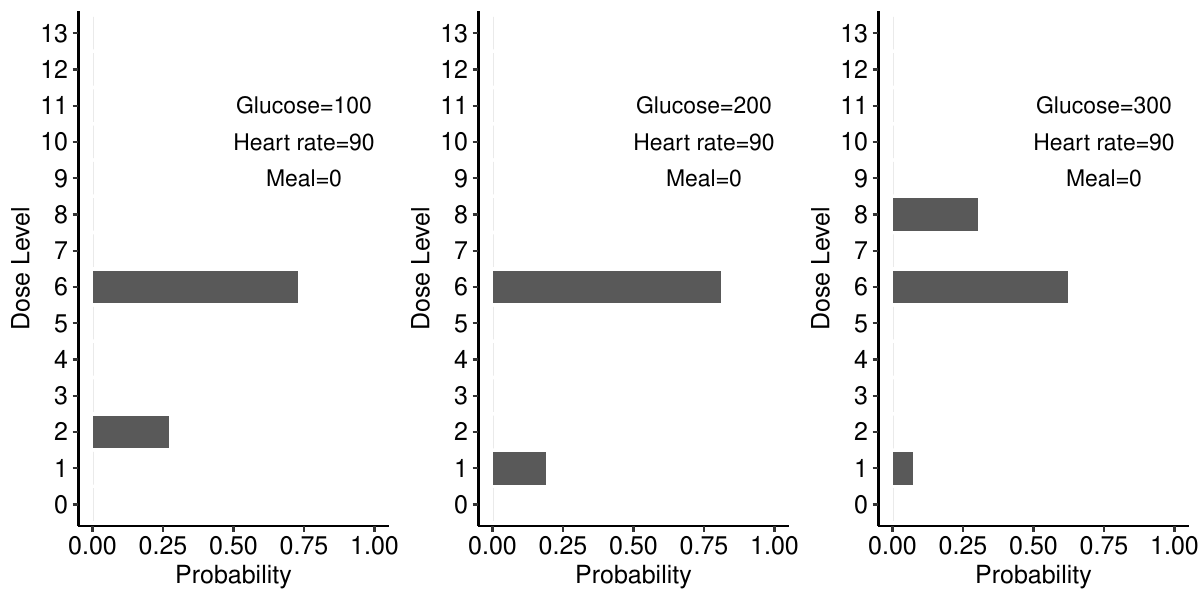}\
\caption{Induced treatment policy for the 1st patient in the 1st cohort given that there is no food intake and the heart rate is normal.}
\label{result1}
\end{figure}

\begin{figure}[htp]
\centering
\includegraphics[scale=0.7]{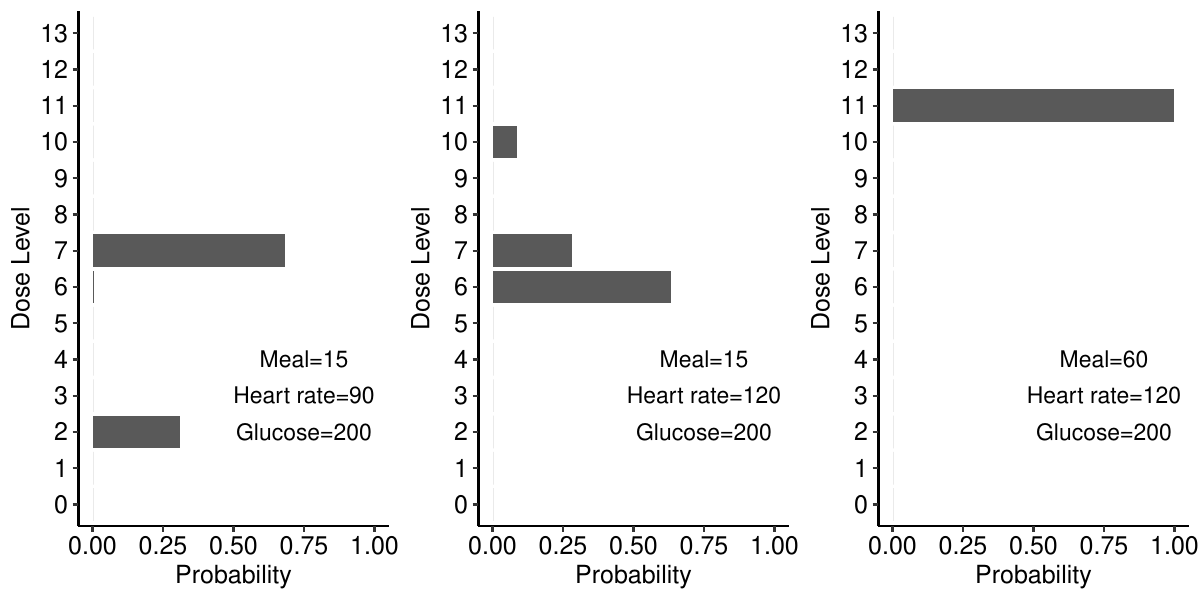}
\caption{Induced treatment policy for 1st patient in the 1st cohort given that glucose level is slightly above average, while there are food intake and the heart rate may also be above average. }
\label{result2}
\end{figure}

To visualize the results, We take the 1st patient in the first cohort as an example. Figure \ref{result1} and Figure \ref{result2} show the induced treatment policy given a new state for that patient. From Figure \ref{result1}, we can see that when there is no food intake and the heart rate is normal (90 bpm), the higher the glucose level for the patient, a higher dose of insulin injection would be recommended. Specifically, when the glucose level is 100 mg/dL, which is still in the normal range, the recommended dose level would be relatively small. When the glucose level is slightly higher than the normal range (200 mg/dL),  the first and 6th dose levels are recommended, while the 6th level has a higher probability compared with the case when the glucose level is 100. When the glucose level is well above the normal range (300mg/dL), the 8th dose level would also be recommended, while the probability of assigning the 1st dose level is significantly reduced.


Note that in Figure \ref{result1}, the highest dose level recommended is the 8th level when there is no food intake, and the heart rate is normal. However, in Figure \ref{result2}, a higher dosage may be recommended when the patient is having meals and the heart rate is above average. For instance, comparing the left and middle panels in Figure \ref{result2}, it seems that a higher heart rate will increase the recommended dose level when the glucose level and carbohydrate intake (15 representing a snack) are the same. Additionally, from the right panel of Figure \ref{result2}, when the heart rate and carbohydrate intakes (60 representing a regular meal) are both above average, an even higher dose of insulin injection would be recommended.

In summary, the recommended treatment policy we get from the algorithm coincides with what we would expect for a patient with diabetes. A higher insulin injection dose would be recommended when the glucose level is higher. At the same time, a higher heart rate and carbohydrate intake could also lead to a higher dose of insulin injection. Such results show our proposed method is practicable and has a clear benefit when applied to real-world settings. Besides, the sparse property of induced policy gives the healthcare provider more flexibility when selecting the treatment.

\begin{figure}[ht]
\centering
\includegraphics[scale=0.7]{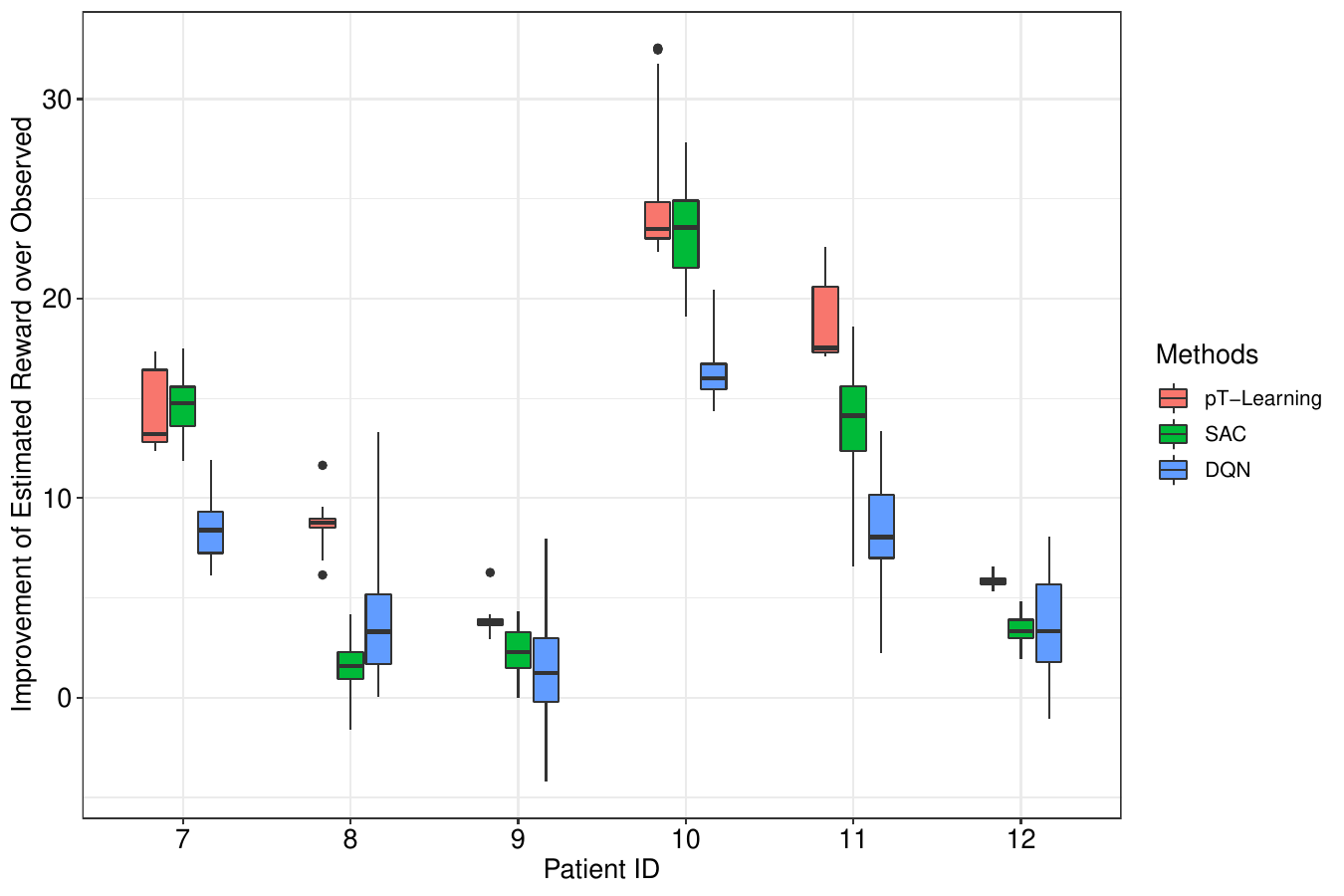}
\caption{Boxplots of improvements on Monte Carlo discounted sum of rewards relative to the baseline of observed discounted sum of rewards. The results included the six patients in the second cohort, and are summarized over 50 simulations with $\gamma=0.9$. }
\label{result3}
\end{figure}

We also compare pT-Learning with two other popular methods in reinforcement learning: Deep-Q network (DQN) \citep{mnih2016asynchronous} and Soft actor-critic (SAC) \citep{christodoulou2019soft}. DQN approximates the Q-function with a neural network and adopts the TD learning framework to solve the optimal policy. The induced deterministic optimal policy would select the action with the largest state-action value at each decision point. Meanwhile, SAC uses a separate network to model the optimal policy besides Q-function and further adds the Shannon entropy of the optimal policy as a bonus term to the V-function. Therefore, SAC would induce an optimal stochastic policy and is directly comparable to pT-learning by using a proximity function $d(x)=\frac{1}{2}x(1-x)$ to the V-function. 

However, we should mention that both DQN and SAC use a neural network to approximate the Q-function, hence needing a large amount of data for model training, which is hardly available in medical settings. For this case study, we use a shallow two-layer perceptron architecture with two linear layers with 64 hidden units to approximate the Q-function for both DQN and SAC. But these methods may not be suitable for a small dataset and do not always have desired statistical properties such as guaranteed convergence and consistency.

Since the data-generating process is unknown, we follow \cite{luckett2020estimating} to utilize the Monte Carlo approximation of the expected discounted sum of rewards to evaluate the model performance. Specifically, we compare the value of $\hat V(S^0_i)$ of each method, where $S_i^0$ denotes the initial state of $i$th trajectory. For DQN and SAC, $\hat V(\cdot)$ refers to the estimated value function. For pT-learning, we consider the lower bound of the estimated value function i.e. $\hat V_{\hat \lambda}(\cdot)-(1-\gamma)^{-1}\hat\lambda\phi(0)$, to mitigate the effect of the sparsity parameter $\lambda$ on the value function. The observed discounted sum of rewards is used as the baseline for each patient. To better evaluate the stability and performance of each method, we randomly select 10 or 15 trajectories from each patient based on available trajectories 50 times and apply these three methods to the selected data.

The boxplots of the improvements on the six patients in the second cohort are presented in Figure \ref{result3}. pT-learning generally has the best performance across all patients. As DQN uses the semi-gradient method, it may diverge under off-policy settings, which explains its largest variance and poor performance compared with SAC and pT-learning. SAC can match the performance of pT-learning on patients 7 and 10 but underperforms on the other four patients. Note that the induced policy in SAC follows a Boltzmann distribution, in which non-negligible probability would be assigned to all actions. This may lead to its unstable performance, especially when the cardinality of the action space is large. On the contrary, pT-learning enjoys the best and most stable performance compared with other methods. This shows that the flexibility of pT-learning to choose deterministic and stochastic policy adaptively has a clear benefit over other methods, especially when a large number of treatment options are available.

\section{Online Experiment and Policy Learning}\label{sec:online}

This section will review some popular adaptive clinical trial designs and online extensions of current policy learning methods. The adaptive clinical trial is distinguished from traditional randomized or observational studies, although they are usually in a finite-horizon setting. It allows the treatment policy to be updated according to historical information during an ongoing clinical trial. Besides, the ethical and sample efficiency concerns also promote the investigation of developing adaptive clinical trials via online policy updating \citep{hu2006theory,thall2013using, chow2014adaptive, hu2015unified}. For example,  \cite{thall2013using} proposed a Bayesian framework to design a trial with multiple combined phases and learn the optimal online policy.

Additionally, their proposed method considers both the sample efficiency and drug toxicity. The limitation is that this method only estimates the average treatment effect and thus fails to account for the patient's heterogeneity information. In contrast, \citep{kim2011battle,renfro2016clinical} considered the subgroup treatment effects via enrichments, but their goal is not estimating the optimal policy or online updating of the clinical trial. Designing more advanced adaptive clinical trials would be helpful for practitioners.

On the other hand, one of the most popular online learning frameworks for an infinite-horizon setting is the $\varepsilon$-greedy strategy. This is developed based on incremental information and tends to balance the exploration and exploitation \citep{lattimore2020bandit}. In detail, the $\varepsilon$-greedy strategy assigns the treatment arm to patients following the estimated policy with the probability $1-\varepsilon$ but lets the $\varepsilon$ chance to select a randomized arm with probability $\varepsilon$. This $\varepsilon$-greedy strategy can avoid sub-optimal solutions and yield an optimal online decision rule. This usually requires $\varepsilon$ to decay to zero as time progresses. There is a trade-off between the large and small $\varepsilon$. Large $\varepsilon$ injects more randomness to select the treatment arm and thus improves the exploration of the policy learning. Small $\varepsilon$ gives attention to the estimated policy, increasing exploitations and sample efficiency.

In recent literature, \cite{luckett2020estimating} follows the $\varepsilon$-greedy to propose an online V-learning framework for updating the estimated policy during the training process. This can also be extended to adaptive clinical trials to improve outcomes. The basic idea of online V-learning is related to offline V-learning, which we have reviewed in Section \ref{V_GCQ}. In online settings, \cite{luckett2020estimating} propose to update the estimated optimal policy $\hat{\pi}_{n}^{t}$ using the data collected up to time $t$. Then adopt the $\varepsilon$-greedy strategy to assign the treatment arm to balance the exploration and exploitation. Most recently, \cite{chen2021statistical} developed an online and robust ordinary least squares to learn a reward model of different actions given the contextual information and then maximize the long-term reward. Their framework also adapts the $\varepsilon$-greedy paradigm and establishes the asymptotic normality to quantify the uncertainty for the optimal decision rule. We may also be interested in updating the estimated optimal policy with patient-specific data. Briefly, the estimated policy $\widehat{\pi}_{n}^{i}$ can be indexed by the patient $i$, which indicates only trained using the data from the patient $i$. This strategy allows more flexibility for learning policies for each patient. Randomness can also be adopted. Finally, we note that an alternative to the $\varepsilon$-greedy strategy to encourage exploration is the upper confidence bound (UCB) sampling. We refer reader to \cite{auer2002using} for more details.

\section{Discussion}

This chapter reviewed the literature on designing and estimating infinite-horizon dynamic treatment regimes. There are still many unsolved problems. We briefly discuss several potential directions for future research. 


The Markov property is fundamental to policy optimizations and evaluations in a finite horizon setting. However, in a real-life environment, outcomes may depend on decisions made much earlier than the immediate, previous time point. This requires a higher-order chain of the Markov decision process. One of the direct applications is to identify the lag-effect variables and conditional independent window for the validity of the Markov property \citep{tsitsiklis1996feature}. However, the literature is very sparse on how to validate such properties \citep{shi2020does}. Overall, it is important to develop new tools that address violations of the current assumptions in the data-generating process. Moreover, when the Markov decision process is only partially observed, additional techniques could be used \citep{ross2011bayesian, ghavamzadeh2016bayesian}. Many of which were developed under a Bayesian framework. 


Most work on developing personalized treatment decisions has focused on a finite number of treatment options. However, learning the individualized decision rule becomes challenging when the treatment space incorporates continuous treatment options. One example is choosing insulin doses from a continuous domain to control blood glucose levels. When learning the optimal dose rules, we may need to account for non-monotone and nonlinear relationships. One of the potential directions is to design an optimal policy that can make continuous action recommendations. \cite{kennedy2017non} considered a method for estimating the average dose-effect with flexible doubly robust covariates adjustment. But the method is not intended for optimal dose-finding problems. \cite{rich2014simulating}
proposed an adaptive strategy, and more recently \cite{schulz2021doubly} developed a doubly robust estimation approach based on a linear model. \cite{chen2016personalized} proposed to optimize the individualized dosing rule by maximizing a local approximation of the value function, solved under the framework of outcome weighted learning \citep{zhao2012estimating}. Recently, kernel-assisted learning with linear dimension reduction \citep{zhou2021parsimonious, zhu2020kernel} for directly learning the optimal dose show good potential. However, all existing methods are limited to single or finite decision stages. Novel methodology frameworks under the infinite horizon setting are much needed. 


For diseases such as cancer or HIV infection \citep{hieke2015conditional, simoneau2020estimating, xue2021multicategory}, survival data often arise. Developing an optimal policy is critical under the survival analysis framework. The main challenge is that treatment and covariate information from patients could be censored in follow-up stages. Also, the true survival time might be unknown for censored patients. Recent developments of the optimal policy estimation under the survival data framework include \citep{goldberg2012q,zhao2014doubly,zhao2020constructing, xue2021multicategory, hager2018optimal}. However, extending existing approaches to an infinite horizon setting is challenging. 


Typically, the goodness of a policy is measured by its value function. Therefore, it is essential to quantify the uncertainty and make statistical inferences about the policy's value function. Specifically, one may construct a confidence interval for the action-value function and its integrated value concerning a given reference distribution. The existing literature in this direction is rare, including the works of \cite{jiang2016doubly,kallus2019efficiently}. Another recent work \citep{shi2020statistical} discussed a recursively updating method to quantify the uncertainty for the off-policy value function. In addition to constructing confidence intervals, validating the policy's value function and measuring how close the policy's value is to the optimal one are also important research questions. 


Confounding and causality are other essential issues in infinite horizon policy evaluation and optimization. The observed behavior policy could involve existing estimated policies and ad hoc human interventions to maximize the long-term rewards. Besides classical literature on causal inference, some recent methods can deal with unmeasured confounders and evaluate treatment effects. \cite{tchetgen2020introduction} proposed a proximal g-computation algorithm in single-stage and two-stage studies. \cite{shi2020multiply} proposed to learn the average treatment effect (ATE) with double-negative control adjustment. A few works considered reinforcement learning with confounded datasets in the computer science literature. For example, \cite{wang2020provably} considered learning an optimal policy in an episodic confounded MDP setting. \cite{bennett2021off} introduced an optimal balancing algorithm for off-policy evaluation in a confounded MDP without requiring mediators to exist. It can be promising and practically useful to develop methods that address confounding issues of infinite horizon policy learning. 

In precision medicine,  data is usually pre-collected as an offline dataset. It implies that no online interaction with the environment is allowed and thus raises an issue on how to exploit the dataset without further exploration. Due to such a lack of exploration, any algorithm for infinite horizon DTR estimations or offline RL suffers from insufficient coverage of the dataset \citep{fujimoto2019off, wang2020statistical, agarwal2020optimistic}. As a result, an
extrapolation error further propagates through each iteration of the algorithm, leading to 
failure in learning an optimal policy. To tackle this issue, pessimistic methods \citep{laroche2019safe,jaques2019way,yu2020mopo,nair2020awac,yang2020off,siegel2020keep,kidambi2020morel,jin2021pessimism} provide various algorithms that enjoy theoretical or empirical success. We may roughly summarize them into two categories: 1) constrain the policy to avoid visiting the states and actions which are less covered by the dataset; 2) penalize the action-value function on such states and actions. In the future, developing more novel methods in policy evaluation and optimization in the principle of pessimism is also crucial.  

Besides the limitations and challenges mentioned above, mHealth requires development in many other aspects. On the practical side, some risk factors can be challenging to measure with wearable devices, and available biomarkers may only serve as surrogates. For instance, when monitoring the mood change of patients with depression, a mobile device cannot collect information as accurate as questionnaires \citep{furukawa2010assessment}. Missing and censoring are quite standard in the mHealth setting \citep{goldberg2012q}. There are issues regarding privacy \citep{bhuyan2017privacy,sunyaev2015availability}. Improving the acceptance of mHealth technology for patients is also essential for facilitating the communication between patients and doctors \citep{silva2015mobile,garavand2017acceptance}.

\bibliography{DTRbook.bib}  
\bibliographystyle{apalike} 

\end{document}